\documentclass[conference]{IEEEtran}
\IEEEoverridecommandlockouts
\usepackage{cite}
\usepackage{amsmath,amssymb,amsfonts}
\usepackage{algorithmic}
\usepackage{graphicx}
\usepackage{textcomp}
\usepackage{xcolor}
\usepackage{algorithm} %format of the algorithm
\usepackage{algorithmic} %format of the algorithm
\usepackage{graphicx}
\usepackage{float}
\def\BibTeX{{\rm B\kern-.05em{\sc i\kern-.025em b}\kern-.08em
    T\kern-.1667em\lower.7ex\hbox{E}\kern-.125emX}}
\begin{document}

\title{Power-Consumption Outage Challenge in Next-Generation Cellular Networks}

\author{\IEEEauthorblockN{Jing Yang, Yi Zhong and Xiaohu Ge}
\IEEEauthorblockA{\textit{dept. School of Electron. Inf. \& Commun.} \\
\textit{Huazhong University of Science and Technology}\\
Wuhan, China \\
\{yang\_jing, yzhong, xhge\}@hust.edu.cn}
\and
\IEEEauthorblockN{Han-Chieh Chao}
\IEEEauthorblockA{\textit{dept. School of Electrical Engineering} \\
\textit{National Dong Hwa University}\\
Hualien, Taiwan \\
hcc@ndhu.edu.tw}
}

\maketitle

\begin{abstract}
The conventional outage in wireless communication systems is caused by the deterioration of the wireless communication link, i.e., the received signal power is less than the minimum received signal power. Is there a possibility that the outage occurs in wireless communication systems with a good channel state? Based on both communication and heat transfer theories, a power-consumption outage in the wireless communication between millimeter wave (mmWave) massive multiple-input multiple-output (MIMO) base stations (BSs) and smartphones has been modeled and analyzed. Moreover, the total transmission time model with respect to the number of power-consumption outages is derived for mmWave massive MIMO communication systems. Simulation results indicate that the total transmission time is extended by the power-consumption outage, which deteriorates the average transmission rate of mmWave massive MIMO BSs.
\end{abstract}


\section{Introduction}
Gigabit-per-second data rates are supported by millimeter wave (mmWave) massive multiple-input multiple-output (MIMO) base stations (BSs) to transmit a huge amount of data to smartphones \cite{key-1}. Moreover, seamless coverage of cellular networks, with the lowest outage probability, is achieved by deploying ultra-dense networks \cite{key-2,key-3}. Considering the existence of the maximum receiving rate of smartphones in \cite{key-4}, a power-consumption outage, which is a new type of outage, still happens in mmWave massive MIMO communication systems with high data rates and seamless coverage. The power-consumption outage occurring in the wireless communication between mmWave massive MIMO BSs and smartphones is caused by the limited computation capability of smartphones \cite{key-4}. The computation capability of smartphones is not only limited by the semiconductor technologies, producing the microchip in smartphones, but also restricted by the heat dissipation of smartphones \cite{key-4}. Smartphones have to shut off the wireless communication with mmWave massive MIMO BSs, i.e., the power-consumption outage occurs, in two cases. One of the two cases is the huge amount of data processing in smartphones without commensurate computation capability of the chip. The other case is that the surface temperature of smartphones exceeds the maximum temperature $45\ {}^\circ \text{C}$ avoiding the low-temperature burn on the users' skin \cite{key-5}. The power-consumption outage is a new challenge in next-generation cellular networks. It is inescapable to model and analyze the power-consumption outage in wireless communication systems.

The conventional outage in wireless communication systems is described as that the received signal power of receivers is less than the minimum received signal power, which makes the communications blocked \cite{key-6,key-7}. Based on the definition of conventional outage, the outage probability of wireless communication systems was proposed as metrics for cellular networks \cite{key-6,key-7}. Moreover, a huge amount of studies had applied the outage probability to analyze the performance of wireless communication systems. The authors in \cite{key-8} analyzed the outage probability of multi-user massive MIMO communication systems with irregular antenna arrays. Moreover, the ergodic achievable sum-rate of massive MIMO communication systems with irregular antennas arrays was derived by the authors in \cite{key-8}. The outage probability of random cellular networks was derived in \cite{key-9}. Based on the outage probability of random cellular networks, the spatial spectrum and energy efficiency of random cellular networks were analyzed \cite{key-9}. Moreover, the outage of cooperative relay channels was analyzed in \cite{key-10}. In vehicular networks, the outage probability of the vehicular device-to-device networks was derived and used to analyze the performance of the vehicular networks \cite{key-11}.

Although the above studies have widely investigated the conventional outage in wireless communication systems, surprisingly little attention has been devoted to the power-consumption outage in wireless communication systems. Inspired by the gap in knowledge, the power-consumption outage in the wireless communication between mmWave massive MIMO BSs and smartphones is modeled and analyzed by using both communication and heat transfer theories. The number of power-consumption outages and total transmission time are derived for numerical analysis. Moreover, the impacts of the power-consumption outage on wireless communication systems have been simulated.

The rest of this paper is outlined as follows. The system model is described in Section II. The power-consumption outage is analyzed in Section III. Moreover, the total transmission time model with respect to the number of power-consumption outages is derived for mmWave massive MIMO communication systems. Detailed simulation results are shown in Section IV. Finally, conclusions are drawn in Section V.

\section{System Model}
\subsection{Downlink Transmission Model}
\begin{figure}[htbp]
\vspace{-0.3cm}
\setlength{\abovecaptionskip}{0pt}
\setlength{\belowcaptionskip}{10pt}
\centerline{\includegraphics[width=8.5cm]{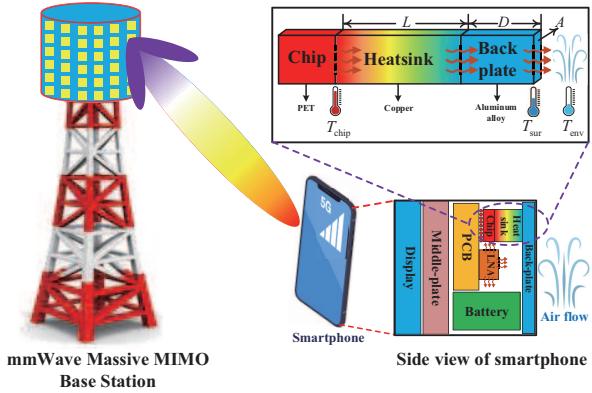}}
\caption{Wireless communication between a mmWave massive MIMO base station and a smartphone.}\label{Fig1}
\label{fig}
\end{figure}
Considering a single cell and one mobile equipment, Fig. 1 illustrates the wireless communication between a mmWave massive MIMO BS and one smartphone. The numbers of antennas in the mmWave massive MIMO BS and the smartphone are configured as ${{M}_{\text{T}}}$ and ${{M}_{\text{R}}}$, respectively. Moreover, ${{M}_{\text{T}}}$ and ${{M}_{\text{R}}}$ satisfy ${{M}_{\text{T}}}\gg {{M}_{\text{R}}}$. The signal-to-noise ratio (SNR) of the $k\text{-th}$ antenna in the smartphone is $\mathrm{SNR}_k$, where $k\in \left\{ 1,\cdots ,{{M}_{\text{R}}} \right\}$. The downlink channel matrix between the mmWave massive MIMO BS and the smartphone is denoted as $\mathbf{H}\in {{\mathbb{C}}^{{{M}_{\text{R}}}\times {{M}_{\text{T}}}}}$. Considering the case that $\mathbf{H}$ is a full-rank matrix, i.e., $\mathrm{rank}\left( \mathbf{H} \right)={{M}_{\text{R}}}$, the downlink rate of the BSs for smartphones is given by
\begin{equation}
{{R}_{\text{downlink}}}=\sum\limits_{i=1}^{{{M}_{\text{R}}}}{B{{\log }_{2}}\left( 1+\mathrm{SNR}_k \right)},
\end{equation}
where $B$ is the bandwidth.

The total number of data transmitted from the mmWave massive MIMO BS to the smartphone is $\Omega $ bits. Moreover, the value of ${{R}_{\text{downlink}}}$ is assumed to be stable in the wireless communication between the mmWave massive MIMO BS to the smartphone. Considering the wireless communication systems without the power-consumption outage, the ideal transmission time of transmitting $\Omega $ bits from the mmWave massive MIMO BS to the smartphone is
\begin{equation}
{{t}_{\text{ideal}}}=\frac{\Omega }{{{R}_{\text{downlink}}}}.
\end{equation}

\subsection{Heat Transfer Model and Power-Consumption Outage}
In Fig. 1, the mass and specific heat of the chip in smartphones are $m$ and ${{c}_{\text{chip}}}$, respectively. The temperature of the chip in smartphones is ${{T}_{\text{chip}}}$. Moreover, the temperature of the smartphone-back-plate, which is near the chip in Fig. 1, is ${{T}_{\text{sur}}}$. During the wireless communication process between mmWave massive MIMO BSs and smartphones, the heat generated by the chip in smartphones raises ${{T}_{\text{chip}}}$, which generates the temperature difference between ${{T}_{\text{chip}}}$ and ${{T}_{\text{sur}}}$. Accordingly, the heat conduction between the chip and the smartphone-back-plate occurs. Firstly, the heat is transferred from the chip to the heat sink whose length and area are $L$ and $A$, respectively. Afterwards, the heat is transferred from the heat sink to the smartphone-back-plate whose thickness is $D$. Considering a copper heat sink, the thermal conductivity of the heat sink is ${{k}_{1}}$ whose value is $\text{401}\ \text{W}/\text{m}\cdot \text{K}$ \cite{key-12}. The material of the smartphone-back-plate is assumed to be 7075-T6 aluminum, which has been widely used in smartphones such as the iPhone 7. Moreover, the thermal conductivity of 7075-T6 aluminum is ${{k}_{2}}$ whose value is $130\ \text{W}/\text{m}\cdot \text{K}$. A local high-temperature area on the smartphone-back-plate occurs when ${{T}_{\text{sur}}}$ is raised by the heat transferred from the chip to the smartphone-back-plate. The environment temperature is ${{T}_{\text{env}}}$, i.e., the temperature of the air. Considering the temperature difference between ${{T}_{\text{sur}}}$ and ${{T}_{\text{env}}}$, the heat accumulated in the smartphone-back-plate is dissipated by free air convection. Moreover, the convection heat transfer coefficient of air is $26.3\ \text{W}/\text{m}\cdot \text{K}$ \cite{key-12}.

The power-consumption outage in smartphones is described as that the heat transferred from the chip to the smartphone-back-plate raises ${{T}_{\text{sur}}}$ to the outage temperature ${{T}_{\text{safe}}}$ \cite{key-4}. ${{T}_{\text{safe}}}=318\ \text{K}$, i.e., $45\ {}^\circ \text{C}$, is the maximum temperature avoiding a low-temperature burn on the users' skin \cite{key-5}. Protection measures in smartphones, e.g., shutting off the wireless communication with mmWave massive MIMO BSs and decreasing the working frequency of the chip in smartphones, have to be taken to reduce the heat generation when ${{T}_{\text{sur}}}$ reaches ${{T}_{\text{safe}}}$. Considering that the wireless communication with mmWave massive MIMO BSs is shut off by smartphones with a certain probability in the case of ${{T}_{\text{sur}}}\ge {{T}_{\text{safe}}}$, the outage probability of the power-consumption outage is
\begin{equation}
{{p}_{\text{outage}}}=\Pr \left( {{T}_{\text{sur}}}\ge {{T}_{\text{safe}}} \right).
\end{equation}
We assume that ${{p}_{\text{outage}}}=1$, i.e., the power-consumption outage occurs in the case of ${{T}_{\text{sur}}}={{T}_{\text{safe}}}$, to simplify the analysis.

In Fig. 1, the heat transferred by the heat sink is divided into two parts, i.e., the heat generated by the chip and the heat transferred from the low noise amplifiers (LNAs) to the chip \cite{key-4}. Therefore, the total heat generation power of the chip in smartphones for the wireless communication process is given by
\begin{equation}
Q_{\text{Total}}^{\text{T}}=Q_{\text{Thermal}}^{\text{chip}}+\lambda {{Q}_{\text{LNA}}},
\end{equation}
where $Q_{\text{Thermal}}^{\text{chip}}$ is the heat generation power of the chip, ${{Q}_{\text{LNA}}}$ is the heat generation power of the LNAs, $\lambda $ is the ratio that the heat transferred from the LNAs to the chip. Considering that each antenna in smartphones is equipped with an LNA, ${{Q}_{\text{LNA}}}$ is calculated as
\begin{equation}
{{Q}_{\text{LNA}}}\text{=}{{M}_{R}}{{P}_{\text{LNA}}}\left( 1-\eta  \right),
\end{equation}
where ${{P}_{\text{LNA}}}$ is the power of one LNA and $\eta $ is the efficiency of LNAs.

The wireless communication between mmWave massive MIMO BSs and smartphones is shut off by smartphones when the power-consumption outage is triggered. Meanwhile, the total heat generation power of the chip is reduced by smartphones. The total heat generation power of the chip in smartphones for the outage process is given by
\begin{equation}
Q_{\text{Total}}^{\text{W}}=Q_{\text{Thermal}}^{\text{chip}}.
\end{equation}

\subsection{Heat Generation Power of the Chip in Smartphones}
Based on Landauer's principle, logically irreversible computations of the chip in smartphones increase the thermodynamic entropy in the environment \cite{key-13}. Considering that there is no mass flow when computations are done by the chip in smartphones, the only way to increase the thermodynamic entropy in the environment is heat transfer \cite{key-14}. Therefore, $Q_{\text{Thermal}}^{\text{chip}}$ is equal to the power of the chip ${{P}_{\text{chip}}}$.

The value of ${{P}_{\text{chip}}}$ belongs to the set $\left\{ P_{\text{chip}}^{\text{T}},P_{\text{chip}}^{\text{W}} \right\}$, where $P_{\text{chip}}^{\text{T}}$ and $P_{\text{chip}}^{\text{W}}$ are the power of the chip in the wireless communication and outage processes, respectively. Moreover, $Q_{\text{Thermal}}^{\text{chip}}$ is given by
\begin{equation}
Q_{{\rm{Thermal}}}^{{\rm{chip}}} = \left\{ {\begin{array}{*{20}{c}}
{P_{{\rm{chip}}}^{\rm{T}}}&{\rm{Communication}\;{\rm{process}}}\\
{P_{{\rm{chip}}}^{\rm{W}}}&{\rm{Outage}\;{\rm{process}}}
\end{array}} \right..
\end{equation}
Considering that both the application processor and the baseband processor have been integrated into the chip of smartphones, $P_{\text{chip}}^{\text{T}}$ consists of the computation power of baseband ${{P}_{\text{BB}}}$ and the system operation power ${{P}_{\text{system}}}$, i.e., $P_{\text{chip}}^{\text{T}}={{P}_{\text{BB}}}+{{P}_{\text{system}}}$. When the power-consumption outage occurs, ${{P}_{\text{chip}}}$ is decreased by smartphones from $P_{\text{chip}}^{\text{T}}$ to $P_{\text{chip}}^{\text{W}}$ to maintain the basic operations of the system in smartphones. Accordingly, $P_{\text{chip}}^{\text{W}}$ is equal to ${{P}_{\text{system}}}$, i.e., $P_{\text{chip}}^{\text{W}}={{P}_{\text{system}}}$.

Based on the computation power model of smartphones in \cite{key-4}, ${{P}_{\text{BB}}}$ is calculated by
\begin{equation}
{{P}_{\text{BB}}}={{R}_{\text{downlink}}}{{K}_{\text{BB}}}{{F}_{0}}\alpha {{E}_{\text{t}}},
\end{equation}
where ${{K}_{\text{BB}}}$ the logic operations per bit of the algorithm in the baseband processor. ${{F}_{0}}$ is the fanout, i.e., the number of loading logic gates. $\alpha $ is the activity factor of transistors for the chip in smartphones. The switching energy consumption of one transistor is ${{E}_{\text{t}}}$ which is
\begin{equation}
{{E}_{\text{t}}}=G{{L}_{\text{bound}}},
\end{equation}
where ${{L}_{\text{bound}}}={{k}_{\text{B}}}{{T}_{\text{env}}}\ln 2$ is the Landauer limit. ${{k}_{\text{B}}}$ is the Boltzmann constant, i.e., $1.38\times {{10}^{-23}}\ \text{J}/\text{K}$. $G$ is the gap between the switch energy consumption for the transistor and the Landauer limit. Considering that 5-nanometer semiconductor technology is used to produce the chip in smartphones, the value of $G$ is estimated to be 454.2 \cite{key-15}.

Considering the initial state of the smartphone in Fig. 1, the power of the chip in smartphones is set to be ${{P}_{\text{system}}}$ for maintaining the basic operations of the system in smartphones. The initial value of ${{T}_{\text{sur}}}$ is $T_{\text{sur}}^{0}$ whose typical value is $T_{\text{sur}}^{0}=303\text{K}$, i.e., $30\ {}^\circ \text{C}$ \cite{key-16}. We assume that the initial state of the smartphone in Fig. 1 is in thermal equilibrium, i.e., the total heat generation power of the chip in smartphones is equal to the heat transfer rate of free air convection. Accordingly, ${{P}_{\text{system}}}$ is estimated by
\begin{equation}
{{P}_{\text{system}}}={{h}_{\text{air}}}A\left( T_{\text{sur}}^{0}-{{T}_{\text{env}}} \right).
\end{equation}

\section{Analyses of Power-Consumption Outage in Wireless Communication Systems}
\subsection{Temperature of Smartphone-Back-Plate and Communication Duration}
A huge amount of data, transmitted by the mmWave massive MIMO BS at a high-data-rate, has to be received and processed by the chip in smartphones, which increases $Q_{\text{Thermal}}^{\text{chip}}$ and ${{Q}_{\text{Total}}}$. When ${{Q}_{\text{Total}}}$ is larger than the heat transfer rate of free air convection, the heat that cannot be timely dissipated by free air convection raises ${{T}_{\text{sur}}}$ until a power-consumption outage occurs in the case of ${{T}_{\text{sur}}}={{T}_{\text{safe}}}$. According to heat transfer theory, the variation of ${{T}_{\text{sur}}}$ for smartphones is a non-equilibrium process, i.e., the temperature varies with time. Based on \textbf{Lemma 1}, the relationship that ${{T}_{\text{sur}}}$ varies with the communication duration $t$ between mmWave massive MIMO BSs and smartphones is obtained.

\vspace{0.1cm}
\textbf{Lemma 1}: Suppose the initial temperature of ${{T}_{\text{sur}}}$ is $T_{\text{sur}}^{\text{start}}$. The equation describing the variation of ${{T}_{\text{sur}}}$ with $t$ satisfies
\begin{equation}
{{T}_{\text{sur}}}=\frac{{{Q}_{\text{Total}}}}{{{h}_{\text{air}}}A}\left( 1-{{e}^{-\frac{zt}{{{c}_{\text{chip}}}m}}} \right)+\left( T_{\text{sur}}^{\text{start}}-{{T}_{\text{env}}} \right){{e}^{-\frac{zt}{{{c}_{\text{chip}}}m}}}+{{T}_{\text{env}}},
\end{equation}
where ${{Q}_{\text{Total}}}\in \left\{ Q_{\text{Total}}^{\text{T}},Q_{\text{Total}}^{\text{W}} \right\}$ is the total heat generation power of the chip in smartphones and $z=1/\left( \frac{L}{{{k}_{1}}A}+\frac{D}{{{k}_{2}}A}+\frac{1}{{{h}_{\text{air}}}A} \right)$.

\vspace{0.1cm}
\textit{Proof}: Considering the one-dimensional steady-state conduction process \cite{key-12}, the heat dissipation power of the chip in smartphones is calculated as
\begin{equation}
q=z\left( {{T}_{\text{chip}}}-{{T}_{\text{env}}} \right),
\end{equation}
where ${{T}_{\text{chip}}}$ is the temperature of the chip in smartphones and $z=1/\left( \frac{L}{{{k}_{1}}A}+\frac{D}{{{k}_{2}}A}+\frac{1}{{{h}_{\text{air}}}A} \right)$.

Applying the energy conservation requirement, the non-equilibrium process that ${{T}_{\text{sur}}}$ varies with the communication duration $t$ between mmWave massive MIMO BSs and smartphones is expressed as a differential equation
\begin{equation}
{{Q}_{\text{Total}}}\cdot dt-q\cdot dt={{c}_{\text{chip}}}m\cdot d{{T}_{\text{chip}}},
\end{equation}
where ${{Q}_{\text{Total}}}\in \left\{ Q_{\text{Total}}^{\text{T}},Q_{\text{Total}}^{\text{W}} \right\}$ is the total heat generation power of the chip in smartphones. Based on the initial condition, i.e., ${{T}_{\text{chip}}}=T_{\text{chip}}^{0}$ in the case of $t=0$, the solution of (13) is derived as
\begin{equation}
{{T}_{\text{chip}}}=\frac{{{Q}_{\text{Total}}}}{z}\left( 1-{{e}^{-\frac{zt}{{{c}_{\text{chip}}}m}}} \right)+\left( T_{\text{chip}}^{0}-{{T}_{\text{env}}} \right){{e}^{-\frac{zt}{{{c}_{\text{chip}}}m}}}+{{T}_{\text{env}}}.
\end{equation}

In the one-dimensional steady-state conduction process, ${{T}_{\text{chip}}}$ and ${{T}_{\text{sur}}}$ satisfy
\begin{equation}
z\left( {{T}_{\text{chip}}}-{{T}_{\text{env}}} \right)={{h}_{\text{air}}}A\left( {{T}_{\text{sur}}}-{{T}_{\text{env}}} \right).
\end{equation}
Therefore, $T_{\text{chip}}^{0}$ and $T_{\text{sur}}^{\text{start}}$ satisfy
\begin{equation}
z\left( T_{\text{chip}}^{0}-{{T}_{\text{env}}} \right)={{h}_{\text{air}}}A\left( T_{\text{sur}}^{\text{start}}-{{T}_{\text{env}}} \right).
\end{equation}
Based on (14), (15) and (16), the equation in (11) can be derived, which completes the proof.

\subsection{Data Receiving Processes in Smartphones}
Considering the wireless communication between mmWave massive MIMO BSs and smartphones at the rate ${{R}_{\text{downlink}}}$, ${{T}_{\text{sur}}}$ is raised by the heat transferred from the chip in smartphones to the smartphone-back-plate. The process raising ${{T}_{\text{sur}}}$ from $T_{\text{sur}}^{0}$ to ${{T}_{\text{safe}}}$ is denoted as the first transmission process ${{S}_{\text{T},1}}$. Based on \textbf{Lemma 1}, the duration of ${{S}_{\text{T},1}}$ is derived as
\begin{equation}
{{t}_{\text{T},1}}=\frac{{{c}_{\text{chip}}}m}{z}\ln \frac{Q_{\text{Total}}^{\text{T}}-{{h}_{\text{air}}}A\left( T_{\text{sur}}^{0}-{{T}_{\text{env}}} \right)}{Q_{\text{Total}}^{\text{T}}-{{h}_{\text{air}}}A\left( {{T}_{\text{safe}}}-{{T}_{\text{env}}} \right)}.
\end{equation}
The number of transmissions between mmWave massive MIMO BSs and smartphones is ${{N}_{\text{T}}}$. When ${{t}_{\text{T},1}}\ge {{t}_{\text{ideal}}}$, transmitting $\Omega $ bits to the smartphone has been finished by the mmWave massive MIMO BS before ${{T}_{\text{sur}}}$ reaching ${{T}_{\text{safe}}}$, which means that the number of transmissions is one, i.e., ${{N}_{\text{T}}}=1$. When ${{t}_{\text{T},1}}<{{t}_{\text{ideal}}}$, transmitting $\Omega $ bits to the smartphone is uncompleted by the mmWave massive MIMO BS with one transmission. Moreover, ${{T}_{\text{sur}}}$ reaches ${{T}_{\text{safe}}}$ when $t={{t}_{\text{T},1}}$, i.e., the power-consumption outage is triggered. Therefore, ${{N}_{\text{T}}}>1$ holds in the case of ${{t}_{\text{T},1}}<{{t}_{\text{ideal}}}$.

After that the power-consumption outage has been triggered, protection measures in smartphones are taken to decrease the total heat generation power of the chip from $Q_{\text{Total}}^{\text{T}}$ to $Q_{\text{Total}}^{\text{W}}$. Moreover, the total heat generation power of the chip is maintained at $Q_{\text{Total}}^{\text{W}}$ by smartphones until that ${{T}_{\text{sur}}}$ is decreased from ${{T}_{\text{safe}}}$ to the end temperature of the power-consumption outage process ${{T}_{\text{wait}}}$, where ${{T}_{\text{env}}}<{{T}_{\text{wait}}}<{{T}_{\text{safe}}}$. The process decreasing ${{T}_{\text{sur}}}$ from ${{T}_{\text{safe}}}$ to ${{T}_{\text{wait}}}$ is denoted as the power-consumption outage process ${{S}_{\text{W}}}$. Based on \textbf{Lemma 1}, the duration of ${{S}_{\text{W}}}$ is calculated as
\begin{equation}
{{t}_{\text{W}}}=\frac{{{c}_{\text{chip}}}m}{z}\ln \frac{Q_{\text{Total}}^{\text{W}}-{{h}_{\text{air}}}A\left( {{T}_{\text{safe}}}-{{T}_{\text{env}}} \right)}{Q_{\text{Total}}^{\text{W}}-{{h}_{\text{air}}}A\left( {{T}_{\text{wait}}}-{{T}_{\text{env}}} \right)}.
\end{equation}
The number of power-consumption outages between mmWave massive MIMO BSs and smartphones is ${{N}_{\text{W}}}$. Moreover, ${{N}_{\text{W}}}$ and ${{N}_{\text{T}}}$ satisfy ${{N}_{\text{W}}}={{N}_{\text{T}}}-1$.

The mmWave massive MIMO BS starts transmitting data to the smartphone with the rate ${{R}_{\text{downlink}}}$ again at the end of ${{S}_{\text{W}}}$. Afterwards, the total heat generation power of the chip in smartphones is growth from $Q_{\text{Total}}^{\text{W}}$ to $Q_{\text{Total}}^{\text{T}}$. Moreover, ${{T}_{\text{sur}}}$ is raised until the power-consumption outage is triggered again. The process raising ${{T}_{\text{sur}}}$ from ${{T}_{\text{wait}}}$ to ${{T}_{\text{safe}}}$ is denoted as the restarted transmission process ${{S}_{\text{T},2}}$. Considering that the initial temperature of ${{T}_{\text{sur}}}$ in ${{S}_{\text{T},2}}$ is ${{T}_{\text{wait}}}$, the duration of ${{S}_{\text{T},2}}$ is derived, based on \textbf{Lemma 1}, as
\begin{equation}
{{t}_{\text{T},2}}=\frac{{{c}_{\text{chip}}}m}{z}\ln \frac{Q_{\text{Total}}^{\text{T}}-{{h}_{\text{air}}}A\left( {{T}_{\text{wait}}}-{{T}_{\text{env}}} \right)}{Q_{\text{Total}}^{\text{T}}-{{h}_{\text{air}}}A\left( {{T}_{\text{safe}}}-{{T}_{\text{env}}} \right)}.
\end{equation}

Considering ${{N}_{\text{T}}}>1$, the data receiving processes in smartphones start from ${{S}_{\text{T},1}}$. Afterwards, the data receiving processes in smartphones are changed back and forth between ${{S}_{\text{W}}}$ and ${{S}_{\text{T},2}}$ until the last transmission. Transmitting $\Omega $ bits to the smartphone is finished by the mmWave massive MIMO BS at the end of the last transmission process. The process of the last transmission between mmWave massive MIMO BSs and smartphones is denoted as ${{S}_{\text{T},3}}$. Although both ${{S}_{\text{T},3}}$ and ${{S}_{\text{T},2}}$ have the same initial temperature of ${{T}_{\text{sur}}}$, i.e., ${{T}_{\text{wait}}}$, the end temperature of the ${{T}_{\text{sur}}}$ in ${{S}_{\text{T},3}}$ satisfies ${{T}_{\text{wait}}}<{{T}_{\text{sur}}}\le {{T}_{\text{safe}}}$. Based on (2), (17) and (19), the duration of ${{S}_{\text{T},3}}$ is given by
\begin{equation}
{{t}_{\text{T},3}}={{t}_{\text{ideal}}}-{{t}_{\text{T},1}}-\left( {{N}_{\text{T}}}-2 \right){{t}_{\text{T},2}},
\end{equation}
where $0\le {{t}_{\text{T},3}}\le {{t}_{\text{T},2}}$. If ${{t}_{\text{T},3}}=0$, the data transmission has been finished at the end of the last ${{S}_{\text{T},2}}$. It is unnecessary to change the data receiving process in smartphones from the last ${{S}_{\text{T},2}}$ to ${{S}_{\text{W}}}$. Accordingly, the last ${{S}_{\text{T},2}}$ is equal to ${{S}_{\text{T},3}}$. If ${{t}_{\text{T},3}}={{t}_{\text{T},2}}$, the data transmission is finished when the end temperature of ${{T}_{\text{sur}}}$ in ${{S}_{\text{T},3}}$ reaches ${{T}_{\text{safe}}}$. Considering the same end temperatures of ${{T}_{\text{sur}}}$ in ${{S}_{\text{T},2}}$ and ${{S}_{\text{T},3}}$, ${{S}_{\text{T},3}}$ is equal to the last ${{S}_{\text{T},2}}$ in the case of ${{t}_{\text{T},3}}={{t}_{\text{T},2}}$. Therefore, ${{S}_{\text{T},3}}$ is equal to the last ${{S}_{\text{T},2}}$ when ${{t}_{\text{T},3}}=0$ or ${{t}_{\text{T},3}}={{t}_{\text{T},2}}$.

\begin{figure}[htbp]
\vspace{-0.3cm}
\setlength{\abovecaptionskip}{0pt}
\setlength{\belowcaptionskip}{10pt}
\centerline{\includegraphics[width=8.5cm]{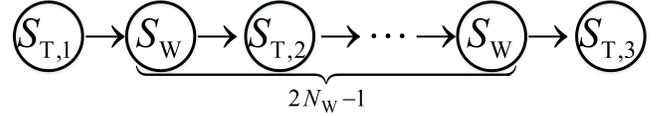}}
\caption{Data receiving processes in smartphones.}\label{Fig2}
\label{fig}
\end{figure}

The data receiving processes in smartphones for the case of ${{t}_{\text{T},1}}<{{t}_{\text{ideal}}}$ is illustrated in Fig. 2. Considering ${{N}_{\text{W}}}={{N}_{\text{T}}}-1$, the total number of ${{S}_{\text{W}}}$ and ${{S}_{\text{T},2}}$ is $2{{N}_{\text{W}}}-1$ in Fig. 2.

\subsection{Number of Power-Consumption Outages and Total Transmission Time}
The data receiving processes in smartphones are divided into four types of processes, i.e., ${{S}_{\text{T},1}}$, ${{S}_{\text{W}}}$, ${{S}_{\text{T},2}}$ and ${{S}_{\text{T},3}}$, due to the power-consumption outage in the wireless communication between mmWave massive MIMO BSs and smartphones. The total transmission time of transmitting $\Omega $ bits from mmWave massive MIMO BSs to smartphones is denoted as ${{t}_{\text{Total}}}$. Moreover, ${{t}_{\text{Total}}}$ is expressed as
\begin{equation}
{{t}_{\text{Total}}}={{t}_{\text{ideal}}}+{{N}_{\text{W}}}{{t}_{\text{W}}},
\end{equation}
where ${{t}_{\text{ideal}}}$ contains the durations of ${{S}_{\text{T},1}}$, ${{S}_{\text{T},2}}$ and ${{S}_{\text{T},3}}$, i.e., ${{t}_{\text{ideal}}}={{t}_{\text{T},1}}+\left( {{N}_{\text{T}}}-2 \right){{t}_{\text{T},2}}+{{t}_{\text{T},3}}$.

To calculate the value of ${{t}_{\text{Total}}}$, the links among $\Omega $, ${{R}_{\text{downlink}}}$, ${{t}_{\text{W}}}$ and ${{N}_{\text{W}}}$ have to be unraveled. Considering ${{t}_{\text{ideal}}}={{t}_{\text{T},1}}+\left( {{N}_{\text{T}}}-2 \right){{t}_{\text{T},2}}+{{t}_{\text{T},3}}$ and $0\le {{t}_{\text{T},3}}\le {{t}_{\text{T},2}}$, ${{N}_{\text{T}}}$ is derived as
\begin{equation}
{{N}_{\text{T}}}=\left\lceil \frac{{{t}_{\text{ideal}}}-{{t}_{\text{T},1}}}{{{t}_{\text{T},2}}} \right\rceil +1,
\end{equation}
where $\left\lceil x \right\rceil $ is the ceiling function of a real number $x$, i.e., outputting the least integer which is greater than or equal to $x$. When ${{t}_{\text{ideal}}}-{{t}_{\text{T},1}}$ is divided exactly by ${{t}_{\text{T},2}}$ in (22), the value of ${{t}_{\text{T},3}}$ satisfies ${{t}_{\text{T},3}}=0$ or ${{t}_{\text{T},3}}={{t}_{\text{T},2}}$, which means that ${{S}_{\text{T},3}}$ is equal to the last ${{S}_{\text{T},2}}$. Considering that the value of ${{t}_{\text{T},2}}$ depends on the value of ${{T}_{\text{wait}}}$ in (19), we assume that the value of ${{T}_{\text{wait}}}$ is chosen appropriately to make that ${{t}_{\text{T},3}}=0$ or ${{t}_{\text{T},3}}={{t}_{\text{T},2}}$, i.e., ${{S}_{\text{T},3}}$ is equal to the last ${{S}_{\text{T},2}}$, to simplify the analysis. Therefore, (22) is transferred to
\begin{equation}
{{N}_{\text{T}}}=\frac{{{t}_{\text{ideal}}}-{{t}_{\text{T},1}}}{{{t}_{\text{T},2}}}+1.
\end{equation}

Based on (18) and (19), the difference value between ${{t}_{\text{W}}}$ and ${{t}_{\text{T},2}}$ is derived as
\begin{equation}
\begin{array}{*{20}{l}}
{\gamma  = {t_{\rm{W}}} - {t_{{\rm{T}},2}}}\\
{\;\;\; = \frac{{{c_{{\rm{chip}}}}m}}{z}\ln \left( {\frac{{Q_{{\rm{Total}}}^{\rm{W}} - {h_{{\rm{air}}}}A\left( {{T_{{\rm{safe}}}} - {T_{{\rm{env}}}}} \right)}}{{Q_{{\rm{Total}}}^{\rm{W}} - {h_{{\rm{air}}}}A\left( {{T_{{\rm{wait}}}} - {T_{{\rm{env}}}}} \right)}}} \right.}\\
{\left. {\;\;\;\;\;\;\;\;\;\;\;\;\;\; \times \frac{{Q_{{\rm{Total}}}^{\rm{T}} - {h_{{\rm{air}}}}A\left( {{T_{{\rm{safe}}}} - {T_{{\rm{env}}}}} \right)}}{{Q_{{\rm{Total}}}^{\rm{T}} - {h_{{\rm{air}}}}A\left( {{T_{{\rm{wait}}}} - {T_{{\rm{env}}}}} \right)}}} \right)}
\end{array}
.
\end{equation}
Moreover, the difference value between ${{t}_{\text{T},1}}$ and ${{t}_{\text{T},2}}$ is calculated, based on (17) and (19), as
\begin{equation}
\begin{aligned}
  & \phi ={{t}_{\text{T},1}}-{{t}_{\text{T},2}} \\
 & \ \ \ =\frac{{{c}_{\text{chip}}}m}{z}\ln \frac{Q_{\text{Total}}^{\text{T}}-{{h}_{\text{air}}}A\left( T_{\text{sur}}^{0}-{{T}_{\text{env}}} \right)}{Q_{\text{Total}}^{\text{T}}-{{h}_{\text{air}}}A\left( {{T}_{\text{wait}}}-{{T}_{\text{env}}} \right)} \\
\end{aligned}
.
\end{equation}
Therefore, the values of $\gamma $ and $\phi $ depend on ${{T}_{\text{wait}}}$.

Based on (2), (23), (24) and (25), the links among $\Omega $, ${{R}_{\text{downlink}}}$, ${{t}_{\text{W}}}$ and ${{N}_{\text{W}}}$ are given by
\begin{equation}
\begin{aligned}
  & \frac{\Omega }{{{R}_{\text{downlink}}}}={{t}_{\text{ideal}}} \\
 & \ \ \ \ \ \ \ \ \ \ \ ={{t}_{\text{T},1}}+\left( {{N}_{\text{T}}}-1 \right){{t}_{\text{T},2}} \\
 & \ \ \ \ \ \ \ \ \ \ \ =\left( {{N}_{\text{W}}}+1 \right){{t}_{\text{T},2}}+\phi  \\
 & \ \ \ \ \ \ \ \ \ \ \ =\left( {{N}_{\text{W}}}+1 \right)\left( {{t}_{\text{W}}}-\gamma  \right)+\phi  \\
\end{aligned}
.
\end{equation}
Therefore, ${{t}_{\text{Total}}}$ is calculated as
\begin{equation}
{{t}_{\text{Total}}}=\frac{\Omega }{{{R}_{\text{downlink}}}}+\frac{{{N}_{\text{W}}}}{{{N}_{\text{W}}}+1}\left( \frac{\Omega }{{{R}_{\text{downlink}}}}-\phi  \right)+\gamma {{N}_{\text{W}}}.
\end{equation}

Based on (27), the average transmission rate of mmWave massive MIMO BSs is ${{R}_{\text{average}}}=\frac{\Omega }{{{t}_{\text{Total}}}}$.

\section{Simulation Results}
Considering the power-consumption outage in the wireless communication between mmWave massive MIMO BSs and smartphones, numerical results are investigated in this section. Moreover, the unit of the temperature in numerical results is converted from Kelvin to centigrade. Other default values of parameters are listed in Table I.

\begin{table}
\vspace{-0.3cm}
\setlength{\abovecaptionskip}{0pt}
\setlength{\belowcaptionskip}{10pt}
\caption{Simulation parameters}

\begin{centering}
\begin{tabular}{|c|c|}
\hline
Parameters & Values\tabularnewline
\hline
Number of antennas in BSs $\ensuremath{M_{\textrm{T}}}$ & 256\tabularnewline
\hline
Number of antennas in smartphones $M_{\textrm{R}}$ & 4\tabularnewline
\hline
Bandwidth $B$ & 1 GHz\tabularnewline
\hline
Mass of the chip $m$ & 2 g\tabularnewline
\hline
Specific heat of the chip $c_{\textrm{chip}}$ & 1030 $\ensuremath{\textrm{J/kg\ensuremath{\cdot}K}}$\tabularnewline
\hline
Length of the heat sink $L$ & 2 mm\tabularnewline
\hline
Area of the heat sink $A$ & 1 $\mathrm{cm}^{2}$\tabularnewline
\hline
Thickness of the smartphone-back-plate $D$ & 1 mm\tabularnewline
\hline
Power of LNA $P_{\mathrm{LNA}}$ & 24.3 mW\tabularnewline
\hline
Efficiency of LNA $\eta$ & 59 \%\tabularnewline
\hline
Ratio of the heat transferred from the LNAs to the chip $\lambda$ & 30 \%\tabularnewline
\hline
Total number of data $\Omega$ & 1 Terabit\tabularnewline
\hline
\end{tabular}
\par\end{centering}
\end{table}

\begin{figure}[htbp]
\vspace{-0.3cm}
\setlength{\abovecaptionskip}{0pt}
\setlength{\belowcaptionskip}{10pt}
\centerline{\includegraphics[width=8.5cm]{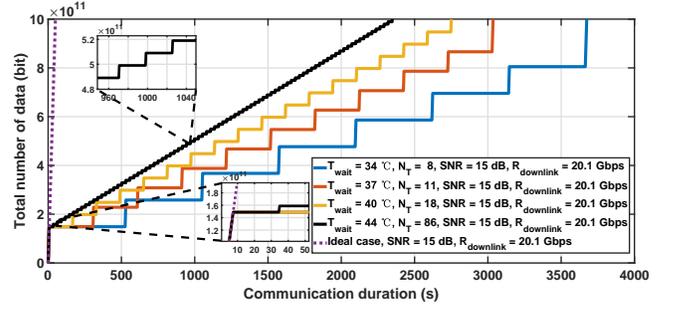}}
\caption{Total number of data as functions of communication duration.}\label{Fig3}
\label{fig}
\end{figure}
Fig. 3 illustrates the total number of data with respect to the communication duration. Moreover, different values of ${{T}_{\text{wait}}}$ have been configured in Fig. 3. The results of Fig. 3 indicate that ${{t}_{\text{Total}}}$ is equal to ${{t}_{\text{ideal}}}$ in the case of $\Omega \le 1.488\times {{10}^{11}}\ \text{bits}$ and the number of transmissions is one, i.e., ${{N}_{\text{T}}}=1$. Moreover, ${{t}_{\text{Total}}}$ is larger than ${{t}_{\text{ideal}}}$ in the case of $\Omega >1.488\times {{10}^{11}}\text{bits}$ and multiple transmissions are needed to transmit $\Omega $ bits from mmWave massive BSs to smartphones, i.e., ${{N}_{\text{T}}}>1$. Based on the results of Fig. 3, the communication duration of the ideal transmission process with ${{R}_{\text{downlink}}}=20.1\ \text{Gbps}$ is 50 seconds when $\Omega =1\times {{10}^{12}}\ \text{bits}$. Considering $\Omega =1\times {{10}^{12}}\ \text{bits}$ and ${{N}_{\text{T}}}>1$, the communication duration in the case of ${{T}_{\text{wait}}}=44\ {}^\circ \text{C}$ is 2,347 seconds, which is 47 times larger than the communication duration of the ideal transmission in Fig. 3. Moreover, the communication duration in the case of ${{T}_{\text{wait}}}=44\ {}^\circ \text{C}$ is the shortest, compared with the communication durations in the other three cases, i.e., ${{T}_{\text{wait}}}=40\ {}^\circ \text{C}$, ${{T}_{\text{wait}}}=37\ {}^\circ \text{C}$ and ${{T}_{\text{wait}}}=34\ {}^\circ \text{C}$. Based on the results of Fig. 3, the communication duration is extended by the power-consumption outage in wireless communication systems.

\begin{figure}[htbp]
\vspace{-0.3cm}
\setlength{\abovecaptionskip}{0pt}
\setlength{\belowcaptionskip}{10pt}
\centerline{\includegraphics[width=8.5cm]{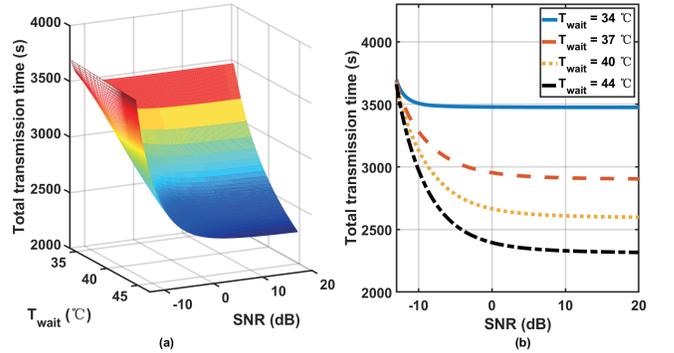}}
\caption{Total transmission time as functions of ${{T}_{\text{wait}}}$ and $\mathrm{SNR}$.}\label{Fig4}
\label{fig}
\end{figure}
The total transmission time with respect to ${{T}_{\text{wait}}}$ and $\mathrm{SNR}$ are shown in Fig. 4. In the results of Fig. 4a, the total transmission time decreases with ${{T}_{\text{wait}}}$. To clearly explain the results in Fig. 4a, four values of ${{T}_{\text{wait}}}$, i.e., $34\ {}^\circ \text{C}$, $37\ {}^\circ \text{C}$, $40\ {}^\circ \text{C}$ and $44\ {}^\circ \text{C}$, are selected and plotted in Fig. 4b. Based on the results of Fig. 4b, the total transmission time decreases with the value of $\mathrm{SNR}$. Considering that a higher value of ${{R}_{\text{downlink}}}$ is corresponding to a larger value of $\mathrm{SNR}$, the total transmission time is still decreased by increasing the value of ${{R}_{\text{downlink}}}$, even the power-consumption outage occurs in wireless communication systems.

\begin{figure}[htbp]
\vspace{-0.3cm}
\setlength{\abovecaptionskip}{0pt}
\setlength{\belowcaptionskip}{10pt}
\centerline{\includegraphics[width=8.5cm]{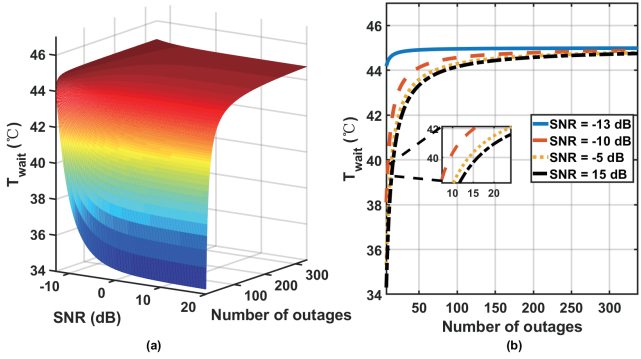}}
\caption{Relationships among ${{T}_{\text{wait}}}$, $\mathrm{SNR}$ and ${{N}_{\text{W}}}$.}\label{Fig5}
\label{fig}
\end{figure}
Fig. 5 illustrates the relationships among ${{T}_{\text{wait}}}$, $\mathrm{SNR}$ and ${{N}_{\text{W}}}$. In the results of Fig. 5a, the value of ${{T}_{\text{wait}}}$ decreases with the value of $\mathrm{SNR}$. To clearly explain the results in Fig. 5a, four values of $\mathrm{SNR}$, i.e., -13 dB, -10 dB, -5 dB and 15 dB, are selected and plotted in Fig. 5b. The results of Fig. 5b indicate that the value of ${{N}_{\text{W}}}$ increases with ${{T}_{\text{wait}}}$. Moreover, the value of ${{N}_{\text{W}}}$ in the case of $\text{SNR}=15\ \text{dB}$ is the maximum, compared with the value of ${{N}_{\text{W}}}$ at the other three cases, i.e., $\mathrm{SNR}=-13\ \text{dB}$, $\mathrm{SNR}=-10\ \text{dB}$ and $\mathrm{SNR}=-5\ \text{dB}$. Based on the results of Fig. 3, Fig. 4 and Fig. 5, the total transmission time decreases with ${{N}_{\text{W}}}$.

\begin{figure}[htbp]
\vspace{-0.3cm}
\setlength{\abovecaptionskip}{0pt}
\setlength{\belowcaptionskip}{10pt}
\centerline{\includegraphics[width=8.5cm]{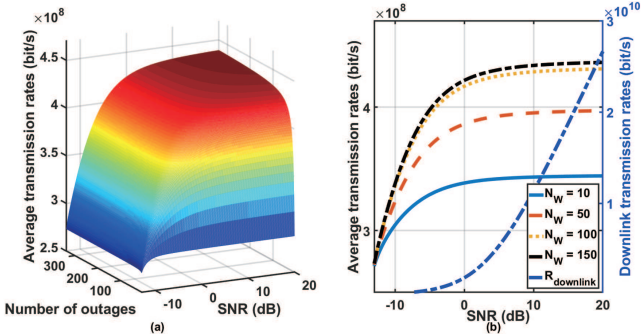}}
\caption{Average transmission rate as functions of ${{N}_{\text{W}}}$ and $\mathrm{SNR}$.}\label{Fig6}
\label{fig}
\end{figure}
The average transmission rate of mmWave massive MIMO BSs with respect to ${{N}_{\text{W}}}$ and $\mathrm{SNR}$ are illustrated in Fig. 6. In the results of Fig. 6a, the average transmission rate increases with ${{N}_{\text{W}}}$. Moreover, the growth rate of the average transmission rate decreases with ${{N}_{\text{W}}}$. In Fig. 6b, the average transmission rate as functions of the value of $\mathrm{SNR}$ are shown. Four different values of ${{N}_{\text{W}}}$, i.e., 10, 50, 100 and 150, have been configured in Fig. 6b. Moreover, the downlink rate of mmWave massive MIMO BSs for smartphones ${{R}_{\text{downlink}}}$ is simulated for comparison. The results in Fig. 6b indicate that the average transmission rate increases with $\mathrm{SNR}$. Moreover, the growth rate of the average transmission rate decrease with $\mathrm{SNR}$. Based on the results of Fig. 6b, the average transmission rate is lower than the value of ${{R}_{\text{downlink}}}$, due to the power-consumption outage in wireless communication systems.

\section{Conclusions}
The power-consumption outage is a new challenge in next-generation cellular networks. Based on both communication and heat transfer theories, this paper has modeled and analyzed the power-consumption outage in the wireless communication between mmWave massive MIMO base stations and smartphones. The total transmission time model with respect to the number of power-consumption outages has been derived for mmWave massive MIMO communication systems. Simulation results indicate that the power-consumption outage happens in wireless communication systems with high data rates due to the limited computation capability of smart devices, e.g., smartphones. Moreover, the power-consumption outage deteriorates the average transmission rate and delays the total transmission time of mmWave massive MIMO BSs. In future work, we plan to analyze the outage in next-generation cellular networks by combing the power-consumption outage and the conventional outage.

\section*{Acknowledgment}
The authors would like to acknowledge the support from the National Key Research and Development Program of China under Grant 2017YFE0121600.

\end{document}